\documentclass{XrU2005}
\usepackage{graphicx}
\usepackage{psfig}

\title{Oxygen and Iron Abundances in Nearby Galaxies Using Ultraluminous X-ray Sources}
\author[1]{Lisa M. Winter}
\author[2]{Richard F. Mushotzky}
\author[1]{Christopher S. Reynolds}

\affil[1]{Department of Astronomy, University of Maryland, College Park, MD, USA 20742}
\affil[2]{Goddard Space Flight Center, Greenbelt, MD, USA 20771}

\begin{document}

\keywords{ISM; ULXs; X-rays}

\maketitle

\begin{abstract}
We determined oxygen and iron abundances of the interstellar medium (ISM) using
K-shell (O) and L-shell (Fe) X-ray photo-ionization edges towards Ultra luminous
X-ray sources (ULXs). We determine the hydrogen column densities (n$_H$) towards
the ULXs from {\it XMM-Newton} archival spectra of 14 ULX sources.  We compare our 
X-ray values with those obtained from radio HI observations for 8 of the sources. 
The X-ray model n$_H$ values are in good agreement with the HI n$_H$ values, 
implying that the hydrogen absorption towards the ULX is not local to the source,
with the exception of M81 X-1.   
Oxygen abundances and iron abundances are roughly solar for
the host galaxies.
\end{abstract}

\section{Introduction}
Within our Milky Way, X-ray binaries have been used as a background through
which to observe the 542\,eV absorption edge produced by photoionization
of the inner K-shell electrons of oxygen \citep{jue04}.  For the first time,
we attempt to extend this type of X-ray absorption study to external galaxies
through the use of ultraluminous X-ray sources (ULXs).

\section{Spectral Fitting}
We analyzed EPIC MOS and PN data from the {\it XMM-Newton} archive for 
14 bright ULX sources.  We required the X-ray spectra to have atleast 
5000\,counts for this study.  A majority of the sources were examined in
Winter, Mushotzky, \& Reynolds (submitted to ApJ), 
where the sources were fitted with simple absorbed blackbody
and power law models.  In this study, we fit all of the sources with a base
model of the {\it grad} model (general relativistic multi-component disk) with
a power law.  We used the \citet{wil00} abundances and absorption models ({\it tbabs}
and {\it tbvarabs}).  The {\it tbabs} absorption model was set to the Galactic value
as obtained from the nH FTOOL in HEASARC, in order to account for Galactic absorption.
We allowed the hydrogen column density, oxygen abundance, and iron abundance to remain
free parameters within the {\it tbvarabs} model.  As in 
Baumgartner \& Mushotzky (submitted to ApJ), we found that
the oxygen absorption values from the three EPIC detectors yielded different values.
Thus, we followed the procedure of Baumgartner \& Mushotzky (submitted to ApJ) 
in adding an {\it edge} model to
account for the differences.  We added an extra edge component to the MOS1 and MOS2
detectors at an energy of 0.53\,keV with optical depths of 0.22 and 0.20 respectively.
The sources and spectral fit parameters will be available in Winter, Mushotzky, \&
Reynolds (in preparation).

\section{Hydrogen Column Densities}
We compared column densities obtained from the X-ray spectral fits with HI column 
densities for Holmberg II X-1, NGC 4559 X-7. NGC 4559 X-10, NGC 5204 X-1 (courtesy 
the WHISP catalog; Swaters et al. 2002) and NGC 247 X-1, M81 X-1, and Holmberg IX X-1 
(Braun 1995).  These are represented as circles in Figure 1.  We also compare our values with 
columns obtained from the E$_{B-V}$ values for M33 X-8 \citet{lon02} and 
M81 X-1 \citep{kon00} using the relationship derived by \citet{pre95}: 
n$_H = 5.3 \times 10^{21}$\,cm$^{-2}$ E$_{B-V}$.  These are represented as triangles.  
The X-ray values 
correspond to the alternate measurements, with the exception of M81 X-1.  For this source 
the X-ray column is greater, indicating the possibility of extra absorption surrounding 
the source.

This result is interesting considering that the X-ray measured column densities are 
along a direct line of sight to the ULXs while the HI measurements are an average over 
a larger beam area.  The agreement between the two measurements implies that the 
ULX sources lie within roughly normal areas of the galaxy (i.e. not in regions of 
higher column density such as a molecular cloud).

\section{O/H ratios}
We compare the oxygen abundance values obtained from the X-ray spectral fits 
with those from studies of HII regions in Figure 2.  The circles are a comparison 
with a study by \citet{pil04} (P-method).  Their method 
was based on spectrophotometric studies of HII regions in the host galaxies.  
We include with the circle symbols the O/H value obtained by \citet{mil95} for 
Holmberg IX X-1 from an optical study of the surrounding HII region as well 
as the O/H value for NGC 1313 (8.4) obtained separately by \citet{cal94} 
and \citet{wal97}.  The triangle symbols represent a comparison with values 
obtained by \citet{gar02} using an alternate method (R23-calibration method).  
Our values are in better agreement with those of \citet{gar02}.

We found that the ratio of Fe/H to O/H obtained through the X-ray spectral modeling 
(tbvarabs) was roughly the solar value.  The iron 
abundance (from the iron L-shell edge located at 851\,keV) was less well-constrained 
than the oxygen abundance.  For the sources with 
greater number of counts ($>$ 20000\,counts) and thus better 
constraints on abundances, the values 
for iron and oxygen do not deviate significantly from solar values.  
The values O/H and Fe/H correspond to: 
	O/H = 12 + log(O$\times$0.00049); 
	Fe/H = 12 + log(Fe$\times2.69\times10^{-5}$), 
where O and Fe are the abundances derived from the spectral fits using 
the Wilms solar values. 

\begin{figure}
\centering
\includegraphics[width=1.0\linewidth]{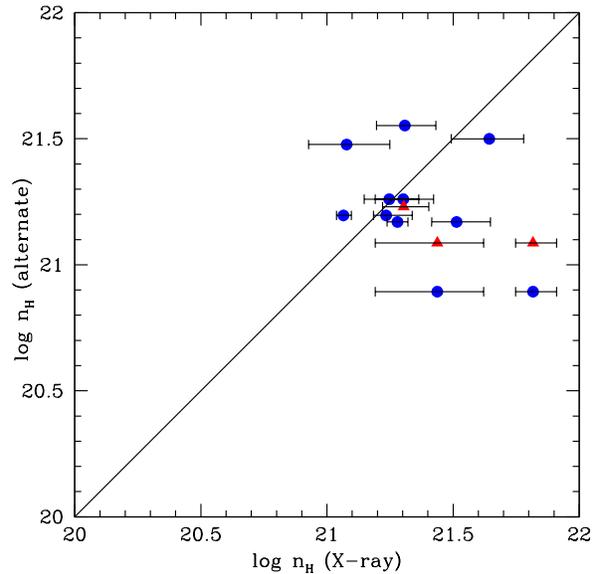}
\caption{Comparison of the hydrogen column density obtained through the X-ray
spectral fits with values from HI radio studies.  The X-ray values are in
good agreement with the radio measurements.  The HI values are an average of
the columns over a larger beam area while the X-ray values are a direct measurement
along the line of sight to the ULX.  This suggests that the sources have average
column densities and therefore do not lie within special areas of the galaxy
(such as a molecular cloud).  The exception is M81 X-1 (the four outlying points)
which may show evidence for extra absorption intrinsic to the source.\label{fig1}}
\end{figure}

\begin{figure}
\centering
\includegraphics[width=1.0\linewidth]{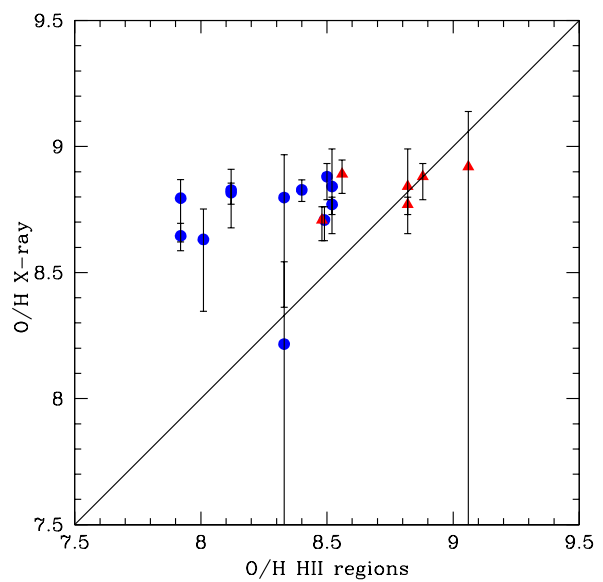}
\caption{Comparison of the X-ray derived O/H ratios with those from optical/UV
studies (see text).  The values are in good agreement with those obtained
by Garnett (2002).  We note that the obtained oxygen abundances are roughly
the solar Wilms values for all of the ULX sources.\label{fig2}}
\end{figure}


\end{document}